\documentclass[10pt]{IEEEtran}
\usepackage{amsmath,amsthm,amssymb}
\usepackage{remark}
\usepackage{boxedminipage}
\usepackage{pgfplotstable}
\usepackage{float}
\usepackage{url}
\def\be{\begin{eqnarray}}
\def\ee{\end{eqnarray}}
\def\bee{\begin{eqnarray*}}
\def\eee{\end{eqnarray*}}
   
\newcommand{\ba}{\begin{align}} 
\newcommand{\ea}{\end{align}}

\usepackage{amsfonts} 
\usepackage{amssymb}

\def\be{\begin{eqnarray}}
\def\ee{\end{eqnarray}}
\def\bee{\begin{eqnarray*}}
\def\eee{\end{eqnarray*}}

\def\E{\mathbb{E}}
\def\P{\mathbb{P}}

\def\E{\mathbb{E}}
\def\P{\mathbb{P}}

\def\bmx{\begin{pmatrix}}
\def\emx{\end{pmatrix}}
\newremark{remark}{Remark}

\def\bmx{\begin{pmatrix}}
\def\emx{\end{pmatrix}}
\begin{document}

\large

\title{Distributed Ledger Technology for Smart Mobility: Variable Delay Models}

\author{A. Cullen\thanks{Andrew Cullen, Pietro Ferraro and Robert Shorten are with the School of Electrical and Electronic Engineering of the
    University College Dublin, Belfield, Ireland.}, P.~Ferraro,
  C.~King,\thanks{Christopher King is with the Department of Mathematics, Northeastern University, Boston, MA 02115 USA.} and
  R.~Shorten}%
  
  \maketitle
  
  \begin{abstract}
Recently, Directed Acyclic Graph (DAG) based Distributed Ledgers have been proposed for various applications in the smart mobility domain \cite{oldpaper}. While many application studies have been described in the literature, an open problem in the DLT community concerns the lack of mathematical models describing their behaviour, and their validation. Building on a previous work in \cite{oldpaper}, we present, in this paper, a fluid based approximation for the IOTA Foundation's DAG-based DLT that incorporates varying transaction delays. This extension, namely the inclusion of varying delays, is important for feedback control applications (such as transactive control \cite{Annaswamy}). Extensive simulations are presented to illustrate the efficacy of our approach.

\end{abstract}
\section{Introduction}

 Directed Acyclic Graph (DAG) based Distributed Ledger Technology (DLT) has recently emerged as an attractive approach for distributed ledger applications in the smart mobility domain \cite{oldpaper}. While distributed ledgers based on Blockchain technology are probably the most well known \cite{Nakamoto}-\cite{Zheng}, others have been developed that are of interest in a control theoretic context. The objective of this paper is to analyse one such alternative ledger structure; namely, the IOTA Foundation's DAG-based distributed ledger, known as the Tangle.
 
To date, applications of DLTs have primarily focussed on two main areas: (i) as a peer-to-peer value transfer e.g. cryptocurrency(e.g., see \cite{Puthal}\cite{SpringerChina}); and (ii) for tracking goods and services in a trustworthy manner in complex supply chains(e.g., \cite{Olnes}).  More recently, DLT has been applied in smart city applications, specifically where the issues of social compliance and the enforcement of social contracts are at the forefront \cite{oldpaper}\cite{sharing1} (e.g. discouraging traffic from
breaking regulations, parking for a limited amount of time in restricted areas, etc.).  For these types of applications DLT is seen as an interesting  enabling mechanism for a number of reasons. First, some DLTs (the IOTA Tangle \cite{Popov}, among others\protect\footnote{e.g. Legicash (\url{https://legi.cash/}) and Byteball (\url{https://byteball.org/})}), specifically designed for high frequency micro-trading, may be more suitable than systems like PayPal, Visa, and even some DLTs like Bitcoin in which vendors will sometimes not process low value transactions.
Second, systems like PayPal, Visa, and many DLTs, may require a transaction fee, thus making their use in digital deposit based systems questionable; namely, where the entire value of the token is intended to be returned to a compliant agent.
Third, in principle, DLT tokens are pseudo-anonymous\protect\footnote{https://laurencetennant.com/papers/anonymity-iota.pdf} due to the use of encrypted adresses. Card based transactions always leave a trail of what was done and when, and are uniquely associated with an individual, the time and location of the spend, and the transaction item. Thus from a privacy perspective (re. Cambridge Analytica and Facebook\protect\footnote{https://www.bbc.com/news/topics/c81zyn0888lt/facebook-cambridge-analytica-data-scandal}), the use of DLT is more like cash and is hence more satisfactory than traditional digital transactions.

 Our interest stems from the possible applications of DLTs in a smart mobility environment, using the digital tokens as a way to enforce the desired level of compliance in the resource-sharing interactions between humans and machines. In this perspective, among the various structures of DLTs, we have chosen to investigate the IOTA Tangle \cite{Popov} for a variety of reasons: first, that there are no explicit transaction fees associated with IOTA, second, the Tangle is designed to support high-frequency micro transactions which is a favourable characteristic in an IoT domain such as the smart mobility one. Finally the structure of the Tangle is well suited for a mathematical representation in which its stability properties can be extrapolated and analysed.

 While many application studies have been described, a significant deficit in the DLT literature concerns the lack of mathematical models describing their behaviour. The Tangle is a DLT in which consesus is achieved through a cooperative mechanism (as opposed to the Blockchain, where users are incentivized to compete through economical fees): newly arrived transactions select two previous transactions to validate (this process will be described more thoroughly in the next Section). The validation procedure involves a proof of work (PoW), that can be modeled as an intrinsic delay mechanism in the dynamics of the system. In our prior work \cite{oldpaper},\cite{TACPaper}, we derived a delayed differential equation model for the Tangle, assuming that the delay was the same for each user and each transaction. Obviously, this is not a realistic hypothesis as different devices and users have different amount of computational power that, in turn, affects how much time the PoW is going to take for each transaction and the nature of the PoW problem results in intrinsic randomness. Accordingly, in this paper we relax this assumption, by accounting for variable delays, and we show, through simulations, that the theoretical predictions are consistent with the behaviour of the Tangle.

 The rest of the paper is organised as follows: In Section \ref{sec: preamble} we provide some examples of smart mobility systems in which DLT can be used to orchestrate human-machine interactions. In Section \ref{sec: Tangle} we describe from an intuitive and high level perspective, the functioning principles of the Tangle. On the basis of this description, in Section \ref{sec: variable delays} we provide the aforementioned mathematical model and the simulations to validate it. Finally, Section \ref{sec: conclusions} summarizes the results of this work and presents future line of research on this topic.

\section{Preamble: Smart Mobility}\label{sec: preamble}

Our starting point in this paper is our interest in smart mobility applications. Generally speaking, we are particularly interested in a class of problems where social compliance to a set of rules is of interest. To provide some context we now give examples of problems that are of interest to us. The interested reader can refer to \cite{oldpaper}, where the following examples are described in detail.

{\em(i) Charge point anxiety :} Many issues impeding the adoption of electric vehicles, such as long charging times, and range anxiety, have been addressed by advances in technology. One human behaviour-related issue remains however.  That is the issue of {\em Charge Point Anxiety}. More specifically, 
public charge points are often occupied by electric vehicle (EV) owners parking there for the entire workday (despite the EV being fully charged). In either scenario the charge point is unavailable to other users  resulting in under-utilization of valuable infrastructure. One solution idea is to develop an adapter to extend the reach of charge points and so allow multiple EVs to connect simultaneously. To this end we designed an adapter \cite{EVAnxiety} which can be connected to a charge point in a `daisy-chained' or `cascaded' manner as shown below in Figure 1. 

\begin{figure}[H]
        \begin{center}
                {\includegraphics[width=3.5in]{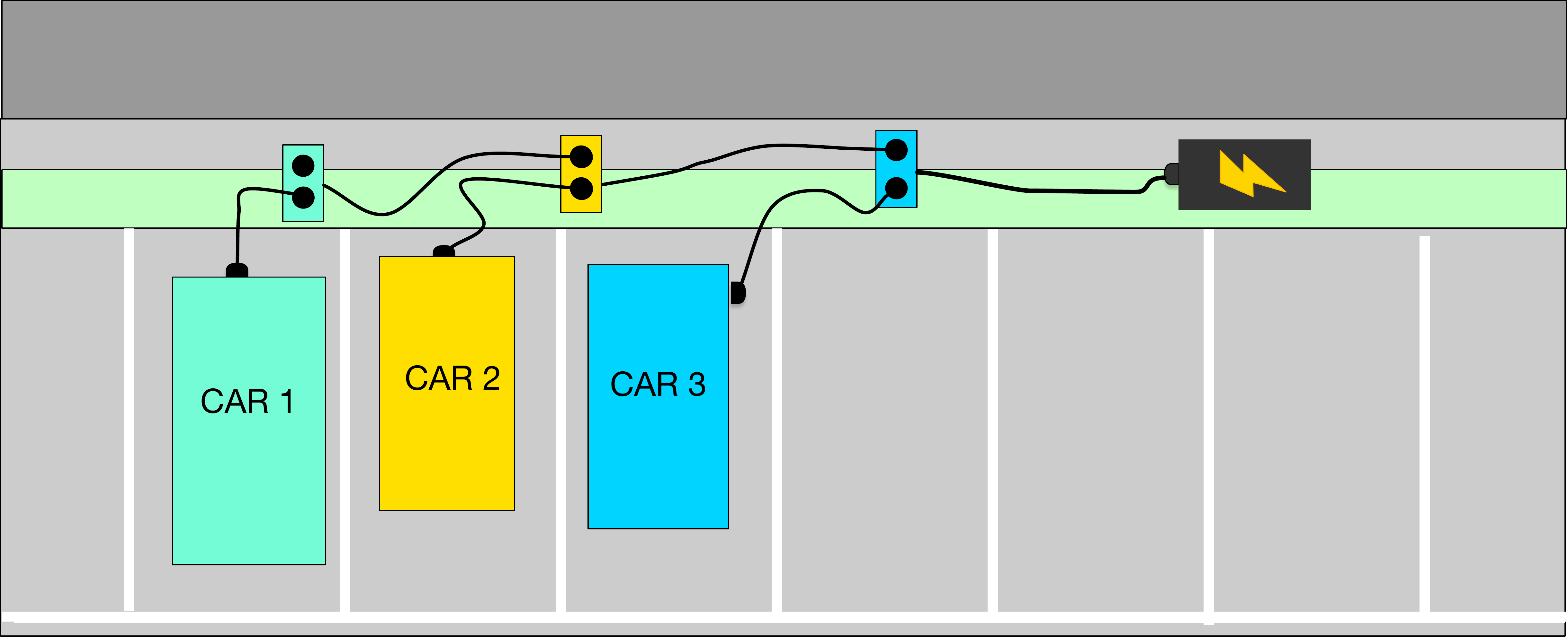}}
                \label{fig:chain}
                \caption{Three vehicles connected to a single charge point and charging simultaneously using three dockChain adapters}
        \end{center}
\end{figure}

To see how DLT technology arises in the context of this example, suppose now that the middle car wishes to disconnect from the chain.  One must hope that the owner of the middle car will connect the green and blue boxes together upon departure (assuming that the cables reach). There is, of course, no guarantee  that this happens. A possible solution  is to incentivise good behaviour using DLTs. That is, to use a digital coin, as part of a deposit system. When the middle car  removes itself from the chain, it deposits a digital token into the charge point; once it reconnects the chain, its token is returned.

{\em (ii) Smart charging hubs for electric bikes :} Pedelecs (e-bikes) are being increasingly viewed as a important part of the e-mobility story. In this context, the opportunity to develop services, for and from such bikes is very appealing, and it is in this context that we are developing, jointly with MOIXA\protect\footnote{http://www.moixa.com/}, a smart battery unit. Roughly speaking, our smart battery is a unit that aggregates the batteries from a number of e-bikes to power ancillary services in a building. In our system, apartment owners would give residents access to an e-bike, and residents would purchase crypto-tokens (whose value would exceed that of a battery) and use a digital deposit system as described above; namely in order to release a battery, users would deposit a token into the system, which would then be returned when the battery is returned.

{\em (iii) Tokenised traffic management :}  A generalization of the two aforementioned examples is the class of \emph{Social Compliance} problems, where the task is to enforce compliance, from agents in a network, to a certain set of rules. In \cite{oldpaper} the authors formalised this class of problems using a network of streets with traffic lights at the intersections, and a population of vehicles, as an example. The compliance goal is that agents (cyclists, drivers) will be `good citizens' and will respect the instructions from the traffic lights. Dynamically-priced tokens are deposited at the lights to control the behaviour of the agents. If the agent obeys the traffic lights when crossing the intersection then the tokens will be returned; otherwise the tokens will be kept by the system.

\section{The Tangle}
\label{sec: Tangle}
\begin{figure}
\includegraphics[width=0.8\columnwidth]{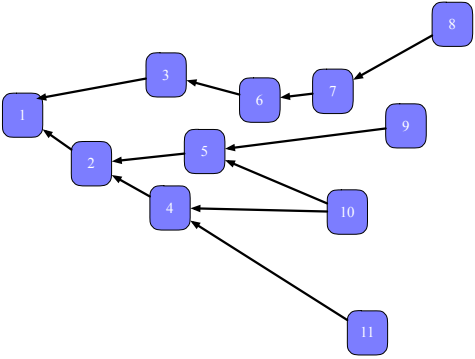}
\caption{Example of a DAG with 11 vertices and 10 edges. All the edges are directed and it is impossible to find a path that connects any vertex with itself.}
\label{Fig: DAG}
\end{figure}

In this paper we are interested in a particular DLT architecture that makes use of DAGs to achieve consensus about the shared ledger. A DAG is a finite connected directed graph with no directed cycles. An example of a DAG is depicted in Figure \ref{Fig: DAG}. The Tangle is a particular instance of a DAG-based DLT \cite{Popov}, where each vertex or \emph{site} represents a transaction  (we will use interchangeably the terms site, transaction and vertex), and 
where the graph, with its topology, represents the ledger. Before being added to the tangle, a new vertex must first approve $m$ (normally two) previous transactions. All yet unapproved sites are called \emph{tips} and the set of all unapproved transactions is called the \emph{tips set}. New transactions will select sites from the tips set for approval (they are not obliged to do so but it is reasonable to expect them to). Each successful approval is represented by an edge of the graph.  The first transaction in the Tangle is called the \emph{genesis} site---the transaction where all the tokens were sent from the original account to all the other accounts---and all transactions either directly or indirectly approve it. Furthermore, in order to prevent malicious users from spamming the network, the approval step requires a Proof of Work (PoW). This step is less computationally intense than its Blockchain counterpart  \cite{Blockchain Security 1} - \cite{Blockchain Security 4}, and can be easily carried out by common IoT devices, but nonetheless introduces some delay for new transactions before they are added to the Tangle. In what follows, we assume that there is a simple way to verify whether the tips selected for approval by a new transaction are consistent with each other and with all the sites directly or indirectly approved by them.

\begin{figure}
\includegraphics[width=0.7\columnwidth]{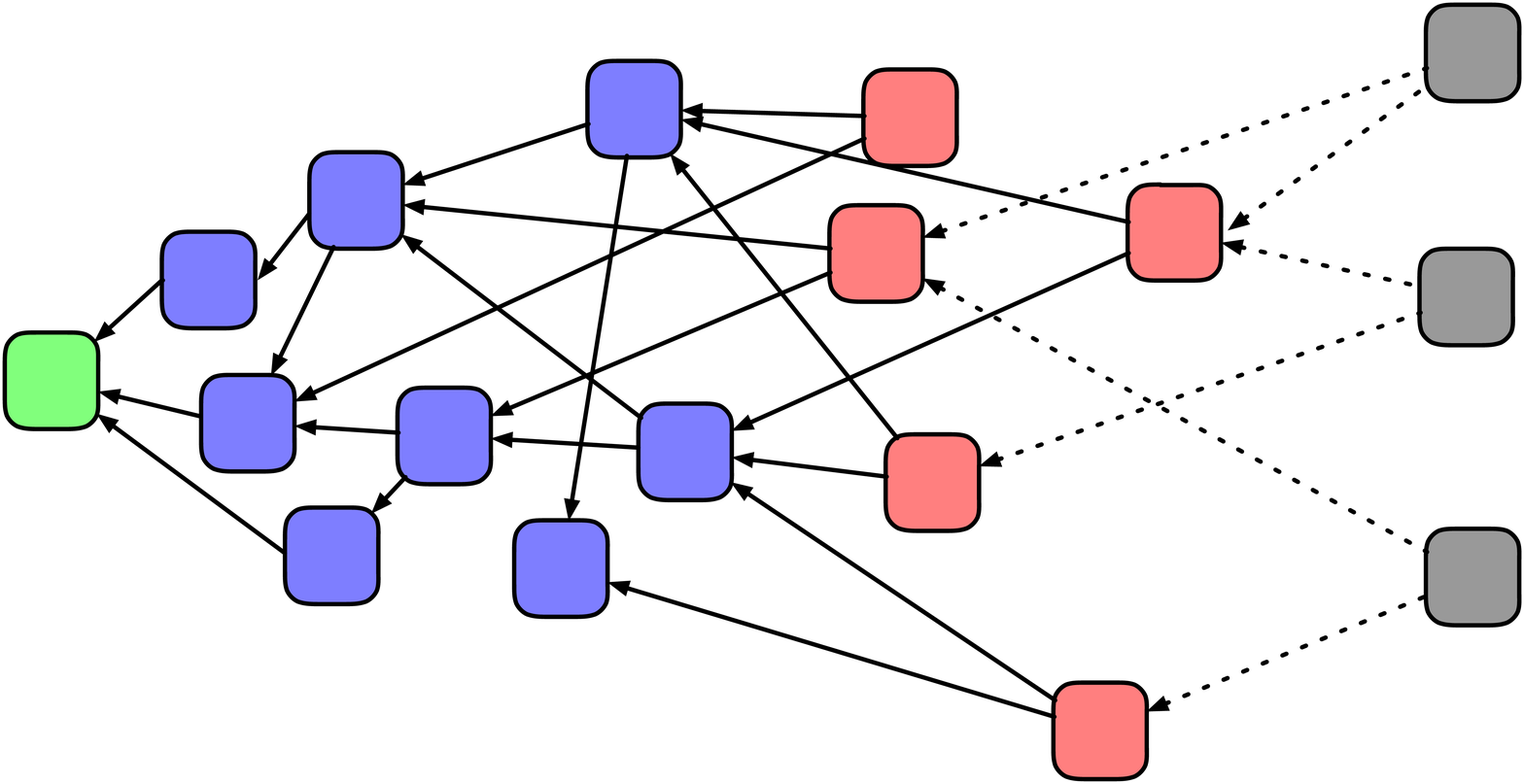}
\includegraphics[width=0.7\columnwidth]{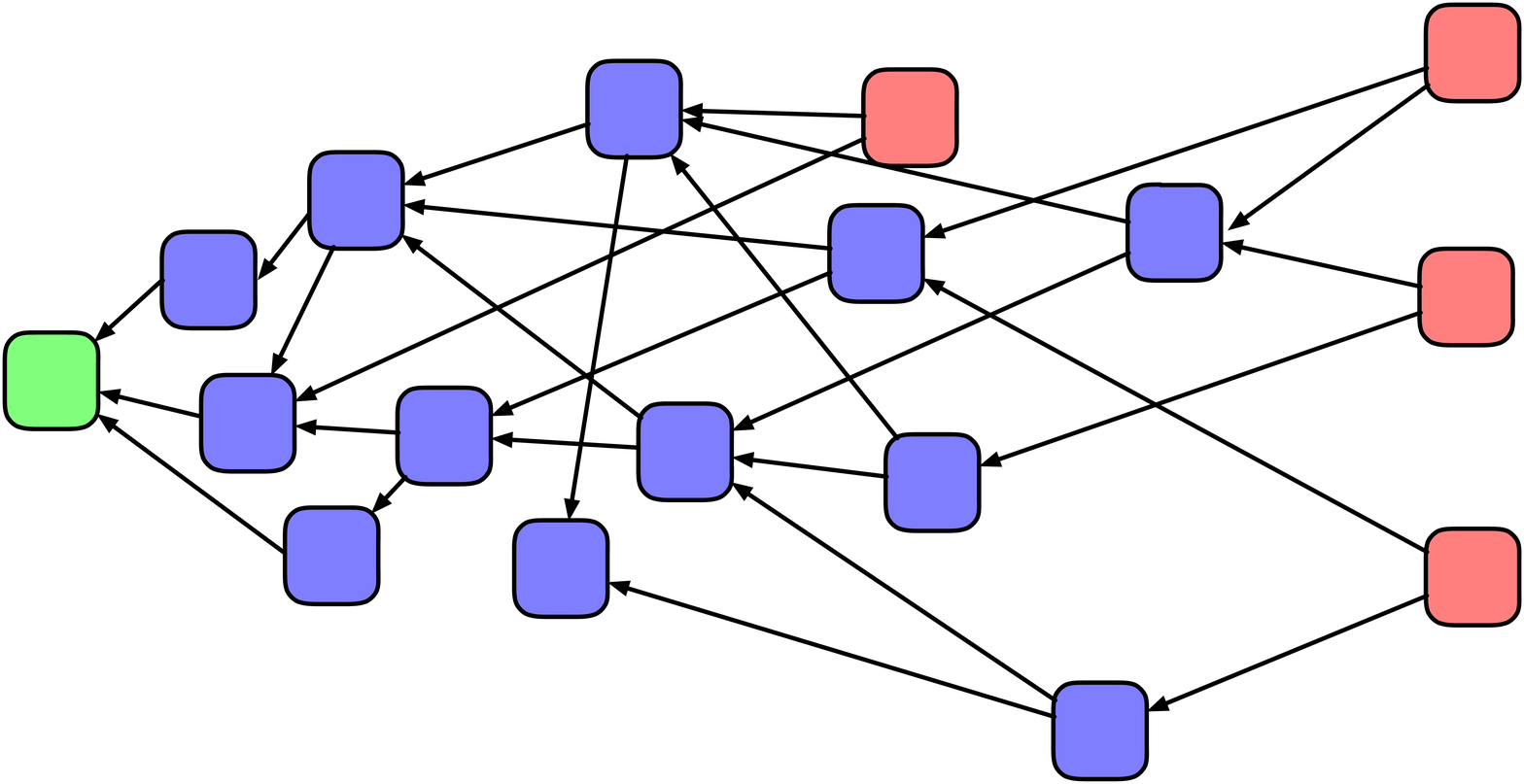}
\caption{Sequence to issue a new transaction. The green site represents the genesis block, the blue sites represent the approved transactions and the red ones represent the tips. The black edges represent approvals, whereas the dashed ones represent transactions that are performing the PoW in order to approve two tips.}
\label{Fig: Tangle}
\end{figure}

As final note, to illustrate the time evolution of the Tangle and visually represent the architecture discussed above, Figure \ref{Fig: Tangle} shows an instance of the Tangle with three new incoming sites (upper panel). The blue blocks are transactions that have already been approved, red blocks represent the current tips of the Tangle and grey blocks are new incoming vertices. Immediately after being  issued,  a new transaction tries to attach itself to two of the network tips. Notice that at this stage, the newly arrived transactions are carrying out the required PoW, and that the tips remain unconfirmed (dashed lines) until this process is over. Once the PoW is finished, the selected tips become confirmed sites and the grey blocks are added to the tips set (lower panel).

\section{ The fluid limit of the tangle with random tip selection and variable delays} \label{sec: variable delays}

\bigskip
The Tangle can be described as an increasing family of DAG's, $\lbrace G(t), t > 0 \rbrace$  where each vertex of $G(t)$ contains the record of a transaction which arrived at or before time $t$. We are interested in describing the growth of the Tangle $G(t)$ in a situation where one central server keeps the record of all the transactions. In a real network there would be multiple local copies of the Tangle, and each user would independently update its own copy. However due to synchronization issues, the presence of multiple independent servers complicates the analysis; therefore, we assume the simpler scenario with one main server, and multiple users accessing it whenever they want to issue a transaction.

We assume that each new transaction selects two tips for approval, and attempts to validate them. If validation fails
the choices are discarded and another
two tips are selected for validation. This continues until the process is successful, and we assume that this whole
validation effort is essentially instantaneous. However after the validation there is a waiting period $H$  during which
the PoW is carried out and the transaction is communicated to the server where the Tangle is stored. 
During this time the approvals of the selected tips are pending, so the tips may still be
available for selection by other new transactions. After the waiting time $H$ the two parent sites
are no longer tips, and so are no longer available for selection by other new transactions (at least, by the ones that follow the protocol)\footnote{It may happen that some of these sites had already ceased  to be tips at an earlier time, due to their being validated by some
other new transaction.}.
In the remainder of this section we assume that the delay times caused by the proof of work for
new transactions are random and 
independent, with some fixed distribution.
Let ${\cal L}(t)$ denote the set of tips at time $t$. Then we assume that when a new transaction arrives at time $t$,
it has selected at random two vertices from the set ${\cal L}(t-H)$ (where $H$ is the random delay time). Thus there are two random
elements in the algorithm; the random choice of delay time $H$, and the random selection of two tips from
the tip set at the earlier time (note that this is only one of the possibile choices for tips selection algorithms \cite{oldpaper}\cite{Popov})

\subsection{Tip selection probability}
Let $\{T_n\}$ denote the increasing sequence of times when new tips are added to the tangle so that
\be
0 \le T_1 \le T_2 \le \cdots \le T_{j} \le \cdots 
\ee
We will label a vertex by the time when it was added to the tangle. Thus vertex $\#j$ was added to the tangle at time $T_j$.
Note that vertex $\#j$ was added as a tip at time $T_j$, and remains in the tip set until some future time when it
is approved. For $j < n$ we define $a_j(T_n)$ to be the indicator variable for the event that vertex $\#j$ is still a tip at time $T_n$.
 That is,
\bee
a_j(T_n) &=& \begin{cases} 1 & \mbox{if $\#j \in {\cal L}(T_n)$ } \\
0 & \mbox{otherwise} \end{cases} \\
a_n(T_n) &=& 1
\eee
Note that $a_j(T_n) \ge a_j(T_m)$ for all $n \le m$. Also recall that $L(t)$ is the number of tips at time $t$, so we have
\be\label{form:L}
L(T_n) = \sum_{j=1}^n a_j(T_n)
\ee

\medskip
Consider now the arrival of a new transaction at time $T_n$.
This new transaction must select two tips for validation from the set ${\cal L}(T_n-H)$, where $H$ is the random
delay time. We define $\tau(T_n)$ to be the set of two vertices which are selected for validation by transaction $\# n$ (i.e., at time $T_n$).
Suppose that vertex $\#j$ is a tip at time $T_n$, and that the random delay time $H$ satisfies $H \le T_n - T_j$.
Since $T_n-H \ge T_j$, this means that the vertex $\#j$ had already been added at time $T_n-H$, and since it is assumed to still be a
tip at time $T_n$, it must also be in the tip set ${\cal L}(T_n-H)$. Thus the probability that vertex $\#j$
is selected for validation at time $T_n$ is simply the probability that any tip is selected for validation
out of all the tips in ${\cal L}(T_n-H)$, which is
\be\label{prob1}
p(T_n - H)= \frac{2}{L(T_n-H)} - \frac{1}{L(T_n-H)^2}
\ee
(the second term in (\ref{prob1}) accounts for the fact that the same tip can be chosen twice by the random tip selection
algorithm). 

\medskip
This result can be formalized in the following way. We define ${\cal F}(n)$ to be the
$\sigma$-algebra generated by the tangle up to time $T_n$. Thus by conditioning on ${\cal F}(n)$
we are fixing the history of the tangle, including of course the tip sets at all previous times.
Also note that $a_j(T_n)=1$ if and only if the
vertex $\#j$ is a tip at time $T_n$. Thus
\be
\P(\#j \in \tau (T_n) \,|\, a_j(T_n)=1, \,\, H, \,\, {\cal F}(n)) =  && \nonumber \\
&& \hskip-2.5in =
\begin{cases} \displaystyle{p(T_n - H)} & \mbox{if $H \le T_n - T_j$} \\
0 & \mbox{if $H > T_n - T_j$}
\end{cases}
\ee
We will write $1_{A}$ to denote the indicator random variable for event $A$; that is
\be
1_{A} = \begin{cases} 1 & \mbox{if $A$ is true} \\
0 & \mbox{if $A$ is false}
\end{cases}
\ee
Then undoing the conditioning on $H$ gives
\be
\P(\#j \in \tau (T_n) \,|\, a_j(T_n)=1, \,\, {\cal F}(n)) =  && \nonumber \\
&& \hskip-2.5in =
\E_{H}\left[1_{{\{H \le T_n - T_j\}}} \, p(T_n - H) \right]
\ee
where the expected value is taken over the distribution of $H$.
We also have 
\be
\P(a_j(T_n)=1 \,|\, {\cal F}(n)) = \E[a_j(T_n) \,|\, {\cal F}(n)]
\ee
and therefore
\be
\P(\#j \in \tau (T_n) \cap {\cal L}(T_n) \,|\, {\cal F}(n) ) =  &&  \\
&& \hskip-1.75in = \E[a_j(T_n) \,|\, {\cal F}(n)] \,\,
\E_{H}\left[1_{\{H \le T_n - T_j\}} \, p(T_n - H)\right]\nonumber
\ee
We now undo the conditioning on ${\cal F}(n)$ (the history of the tangle) and obtain
\be
\P(\#j \in \tau (T_n) \cap {\cal L}(T_n)) = \E\left[a_j(T_n)  \,\,
1_{{\{H \le T_n - T_j\}}} \, p(T_n - H)\right]
\ee
Note also that  $a_j(T_{n+1}) - a_j(T_n) = -1$ if and only if the
vertex $\#j$ is a tip at time $T_n$ and is approved by the new transaction which arrives at time $T_n$,
and is zero otherwise.
Therefore we get
\be\label{eq1}
\E[a_j(T_{n+1}) - a_j(T_n) ] =\nonumber \\
&& \hskip-1.75in = - \E\left[a_j(T_n) \,\,
1_{{\{H \le T_n - T_j\}}} \, p(T_n - H)\right]
\ee

\subsection{The fluid limit}
The fluid limit is reached as the arrival rate $\lambda$ goes to infinity.
For convenience we will assume that the time between arrivals is fixed and equal to $\lambda^{-1}$,
so $T_n = n \lambda^{-1}$, and we extend $a_j$ to be piecewise constant in each interval $[T_n,T_{n+1})$.
Given $s > 0$
let $m = \lfloor \lambda s \rfloor$, and define the set ${\cal A}(s) = \{m,m+1,\dots,m+q\}$ where $q$ is an integer
depending on $\lambda$ such that $\{q \rightarrow \infty$, $q \lambda^{-1} \rightarrow 0\}$
as $\lambda \rightarrow \infty$. We define
\be
b(t,s) = q^{-1} \, \sum_{j \in {\cal A}(s)} a_j(t), \qquad l(t) = \lambda^{-1} \, L(t)
\ee
(assuming that $t \ge s + q \lambda^{-1}$ in $b$).
Our main assumption for the fluid limit is that
$b(t,s)$ and $l(t)$ converge to non-random differentiable functions as $\lambda \rightarrow \infty$.
So in particular we assume that
\be
q^{-1} \, \sum_{j \in {\cal A}(s)} a_j(t)  \rightarrow b(t,s) \quad
\mbox{as $\lambda \rightarrow \infty$}
\ee
We also assume that for all $j \in {\cal A}(s)$ and all $n \in {\cal A}(t)$, as $\lambda \rightarrow \infty$,
\be
q^{-1} \, \sum_{j \in {\cal A}(s)} a_j(T_n) \, 1_{{\{H \le T_n - T_j\}}} \, p(T_n - H)  &&
\nonumber \\
&& \hskip-3in = \lambda^{-1} \, b(t,s) \, 1_{{\{H \le t - s\}}} \, \frac{2}{l(t-H)} + o(\lambda^{-1})
\ee

\medskip
We now sum over $j \in {\cal A}(s)$ and $n \in {\cal A}(t)$ in (\ref{eq1}), leading to
\be
& \hskip-0.2in \sum_{n \in {\cal A}(t)} \, q^{-1} \, \sum_{j \in {\cal A}(s)} \E[a_j(T_{n+1}) - a_j(T_n) ] = \nonumber\\
&  \hskip-1in  =\sum_{n \in {\cal A}(t)} \, b(T_{n+1},s) - b(T_n,s)= \nonumber \\
& \hskip-1.47in = b(t + q \lambda^{-1}, s) - b(t,s)= \nonumber \\
& \hskip-0.1in= - \sum_{n \in {\cal A}(t)} \, \lbrace\lambda^{-1} \,  b(t,s) \, \, \E_{H}\left[1_{\{H \le t -s\}}  \frac{2}{l(t-H)} \right] +  \nonumber \\
& + o(\lambda^{-1})\rbrace  = - q \, \lambda^{-1}  \, b(t,s) \, \, \E_{H}\left[1_{{\{H \le t - s\}}} \, \frac{2}{l(t-H)} \right] +  \nonumber \\
& \hskip-2.81in + o( q \lambda^{-1})
\ee
Since $q \lambda^{-1} \rightarrow 0$ as $\lambda \rightarrow \infty$, we get
\be
\frac{\partial b}{\partial t}(t,s) = - b(t,s) \, \, \E_{H}\left[1_{{\{H \le t - s\}}} \, \frac{2}{l(t-H)} \right]
\ee

It is convenient to change variables at this point, and define the age of the current tips to be $v = t-s$,
and define a new density $g(t,v) = b(t,s)$, where $g(t,v)$ represents the density of tips present at time $t$ which were added at time $t-v$ (or, equivalently, the density of tips present at time $t$ with age $v$). 
Note also the total number of tips (rescaled by $\lambda^{-1}$) is
\be
l(t) = \int_{0}^{t} b(t,s) \, ds = \int_{0}^{t} g(t,v) \, dv
\ee
Furthermore the condition $a_n(T_n)=1$ leads to the condition $b(t,t)=1$, which in turn gives
\be
g(t,0) = 1
\ee

\medskip
In terms of these new variables, the fluid limit is described by the following set of equations:
\be\label{fluid2} 
\frac{\partial g}{\partial t} + \frac{\partial g}{\partial v} &=& - g(t,v) \, \E_{H} \left[ 1_{\displaystyle{\{H \le v\}}} \, \left(\frac{2}{l(t-H)}  \right) \right]  \nonumber\\
l(t) &=& \int_{0}^{t} g(t,v) \, dv  \nonumber \\
g(t,0) &=& 1
\ee

\medskip
The right side of (\ref{fluid2}) can be written more explicitly using the pdf for $H$. That is, assume that $H$ is continuous with pdf
$f(x)$, then  (\ref{fluid2}) can be written
\be
\frac{\partial g}{\partial t} + \frac{\partial g}{\partial v} = - g(t,v) \, \int_{0}^{v} \, \frac{2}{l(t-x)}  \, f(x) \, d x
\ee

\subsection{Comparison with previous work for fixed delay}
In previous work \cite{oldpaper} the case of fixed delay $H=h$ was analyzed by different means, and it was shown that the 
fluid limit was described by a delay differential equation. Here we compare that result with the present work.
For the case of constant $H=h$, the PDE (\ref{fluid2}) leads to the equation
\be\label{fluid2a}
\frac{d l}{d t} = \begin{cases} 1 & t < h \\
\displaystyle{1 - \frac{2}{l(t-h)} \, \int_{h}^t g(t,v) \, dv} & t \ge h \end{cases}
\ee
In \cite{oldpaper} the following equation was derived:
\be\label{fluid2b}
\frac{d l}{d t} = 1 -  \frac{2}{l(t-h)} \, x(t-h)
\ee
where $x(s)$ is the number of `free' tips at time $s$, which is the number of tips that have not yet been selected at time $s$ for validation
by any newly created transactions. Since the validation time is fixed to be $h$, it follows that all of these free tips at time $t-h$ must still be tips
at time $t$. Furthermore any `pending' tips at time $t-h$ will no longer be tips at time $t$. Thus the set of tips at time $t$ will
consist of the free tips at time $t-h$, plus any additional tips that arrived in the time interval $[t-h,t]$. These latter tips are the tips
whose age is less than $h$, thus we can write
\be
l(t) = x(t-h) + \int_{0}^h g(t,v) \, dv
\ee
Combining with the relation $l(t) = \int_{0}^{t} g(t,v) \, dv$ we deduce that for $t \ge h$
\be
x(t-h) = \int_{h}^{t} g(t,v) \, dv
\ee
Therefore the two expressions (\ref{fluid2a}) and (\ref{fluid2b}) are identical for all $t \ge h$.
Hence the method presented here leads to the same result as the method from \cite{oldpaper}.

\subsection{The stationary solution}
We expect that the solution of the system (\ref{fluid2}) will converge to a time-independent solution as
$t \rightarrow \infty$. We can compute this time-independent solution: assume that there is a function
$g(v)$ and constant $l$ such that
\be
g(t,v) \rightarrow g(v), \quad l(t) \rightarrow l \quad \mbox{as $t \rightarrow \infty$}
\ee
Substituting in (\ref{fluid2}) we find
\be\label{ODE1}
g'(v) = - g(v) \, \frac{2}{l} \, \P(H \le v), \quad g(0)=1
\ee
which leads to
\be\label{ODE-sol}
g(v) = \exp\left[- \frac{2}{l} \, \int_{0}^{v} \P(H \le u) \, du\right]
\ee
This can be used to get an implicit equation for $l$:
\be\label{ODE2}
l = \int_{0}^{\infty} g(v) \, d v
\ee

In what follows, we consider some specific examples for delays and derive, for each of them, the corresponding tip equilibrium. We make use of several Monte Carlo simulations of the Tangle to validate our theoretical predictions.

\subsection{Special case: fixed delay $H=h$}
\begin{figure}
\includegraphics[width=1\columnwidth]{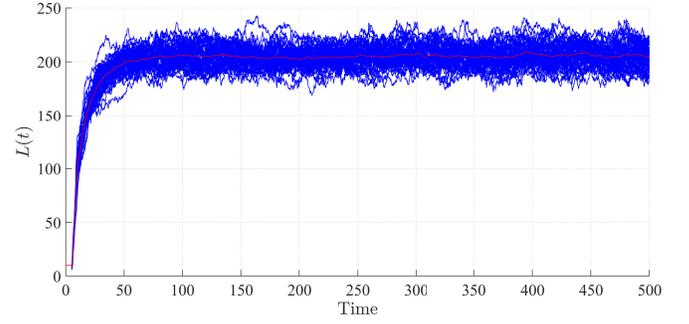}
\caption{150 Monte Carlo simulations of the Tangle with constant delay ($\lambda = 20, h = 5$). The single realizations are shown in blue, while the average value is shown in red. Notice that we obtain the predicted average value $L=200$.}
\label{Fig: Const delayl}
\end{figure}

Here we assume that $H=h$ is constant for all tips. Then solving (\ref{ODE1}) we get
\be
g(v) = \begin{cases} 1 & v \le h \\ e^{-2 (v-h)/l} & v > h \end{cases}
\ee
We can also use (\ref{ODE2}) to compute $l$: this gives
\be
l = \int_{0}^{\infty} g(v) \, d v = h + \int_{0}^{\infty} e^{-2 (v-h)/l} \, dv = h + \frac{l}{2}
\ee
which leads to the value
\be
l = 2 h
\ee
Figure \ref{Fig: Const delayl} shows 150 Monte Carlo simulations of the Tangle with fixed delay ($\lambda = 20, h = 5$). Note that the average value corresponds to $L=\lambda l=2 \lambda h = 200$.

\subsection{Special case: exponential delay $H$}
\begin{figure}
\includegraphics[width=1\columnwidth]{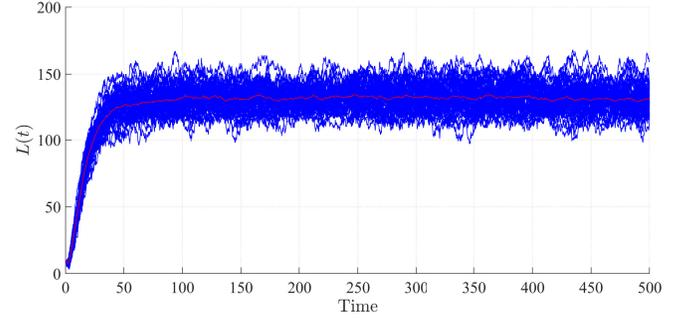}
\caption{150 Monte Carlo simulations of the Tangle with exponential delay ($\lambda = 20, \mu = 0.2 = 5^{-1}= h^{-1}$). The single realizations are shown in blue, while the average value is shown in red. Notice that we obtain the predicted average $L=128$.}
\label{Fig: exponential delayl}
\end{figure}

Here we assume that $H$ is exponential with rate $\mu$, so that (\ref{ODE1}) is
\be
g'(v) = - g(v) \, \frac{2}{l} \, (1 - e^{-\mu v}), \quad g(0)=1
\ee
This leads to the solution
\be
g(v) = \exp\left[ \frac{2}{l} \, \left(v + \mu^{-1} e^{-\mu v} - \mu^{-1}\right)\right]
\ee
Using (\ref{ODE2}) we can compute $l$. With $h = \mu^{-1}$ this gives
\be
l = 1.2839 \, h
\ee
Figure \ref{Fig: exponential delayl} shows 150 Monte Carlo simulations of the Tangle with exponential delay ($\lambda = 20, \mu = 0.2 = 5^{-1}= h^{-1}$). Note that the average value corresponds to $L=\lambda l = 1.28 \,\lambda \, h = 128$.

\subsection{Special case: uniform delay $H$}
\begin{figure}
\includegraphics[width=1\columnwidth]{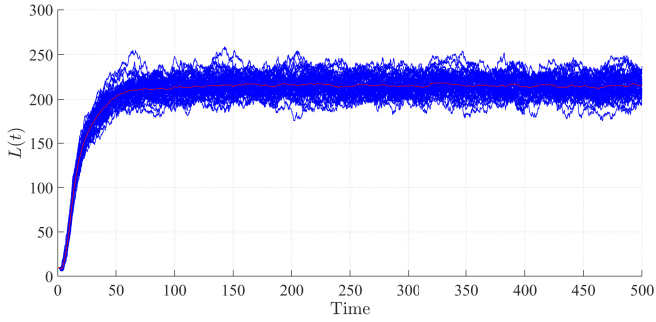}
\caption{150 Monte Carlo simulations of the Tangle with uniform delay ($\lambda = 20, h_0=1$, $h_1=11$). The single realizations are shown in blue, while the average value is shown in red. Notice that we obtain the predicted average $L=214$.}
\label{Fig: uniform delayl}
\end{figure}

Here we assume that $H$ is uniform on some interval $[h_0,h_1]$.
This gives
\be
\int_{0}^v \P(H \le u) \, du = \begin{cases} 0 & v \le h_0 \\
& \\
\displaystyle{\frac{(v - h_0)^2}{2 (h_1 - h_0)}} & h_0 \le v \le h_1 \\
& \\
\displaystyle{v - \frac{h_0 + h_1}{2}} & v > h_1 \end{cases}
\ee
Also we define $\beta = \sqrt{h_1-h_0}$ then the equation for $l$ is
\be
l = h_0 + \frac{l}{2} \, e^{-\beta^2/l} + \beta \int_{0}^{\beta} e^{ - w^2/l} \, d w
\ee
One particular case: $h_0=1$, $h_1=11$, gives $l = 10.69$. In terms of the mean delay $h = 6$ this is
\be
l = 1.782 \, h
\ee
Figure \ref{Fig: uniform delayl} shows 150 Monte Carlo simulations of the Tangle with uniform delay ($\lambda = 20, h_0=1$, $h_1=11$). Note that the  average value corresponds to $L=\lambda l= 1.78 \lambda h = 214$.

\section{Conclusions} \label{sec: conclusions}

In this paper we present a model for the Tangle that takes into account variable delays for the Proof of Work. Under the assumption of a high arrival rate, we derive a general fluid model for the Tangle, of which the equations presented in \cite {oldpaper} represent a particular case. Accordingly, we show that under the assumption of constant delays, the new set of equations match the ones in \cite {oldpaper}.

\end{document}